\documentclass[11pt]{article}
\usepackage{times}
\usepackage{soul}
\usepackage{url}
\usepackage[utf8]{inputenc}
\usepackage{graphicx}
\usepackage{amsmath}
\usepackage{booktabs}
\usepackage{algorithm}
\usepackage{algorithmic}
\usepackage{mathrsfs}
\usepackage[T1]{fontenc}
\usepackage{verbatim}
\usepackage{color}
\usepackage{stmaryrd}
\usepackage{enumerate}
\usepackage{xspace}
\usepackage{balance} 
\usepackage{multirow}
\usepackage{adjustbox}
\usepackage{amssymb}
\usepackage[breaklinks]{hyperref}
\usepackage{tikz}
\usepackage{cite}

\usetikzlibrary{decorations,decorations.pathmorphing,arrows,shapes,%
  automata,backgrounds,fit,calc,petri,patterns,matrix}
\usepackage{array}
\newcolumntype{H}{>{\setbox0=\hbox\bgroup}c<{\egroup}@{}}

\usepackage{authblk}

\usepackage{wojtek15logics,wojtek15other}

\newcommand{\Act}{{Act}}
\newcommand{\state}{q}
\renewcommand{\act}{\alpha}
\newcommand{\mov}{\profile{\act}}
\newcommand{\coalition}{C}
\newcommand{\distribution}{\mathsf{d}}
\newcommand{\Dist}{\text{Dist}}
\newcommand{\System}{M}
\newcommand{\legal}{R}
\newcommand{\val}{\Val}
\newcommand{\obsrel}[1][]{\sim_{#1}}

\renewcommand{\path}{\lambda}
\newcommand{\history}{h}
\newcommand{\History}{\text{Hist}}

\newcommand{\last}{\mbox{last}}
\renewcommand{\str}{\sigma}

\newcommand{\setstrat}{\mbox{\emph{Str}}}
\renewcommand{\profile}[1]{\boldsymbol{#1}}

\newcommand{\PCTLs}[1][]{\lan{PCTL_\stratstyle{#1}^*}}
\newcommand{\PATL}[1][]{\lan{PATL_{\stratstyle{#1}}}}

\newcommand{\PATLs}[1][]{\lan{PATL_\stratstyle{#1}^*}}

\newcommand{\ie}         {i.e., }

\renewcommand{\model}{\mathcal{S}}

\newcommand{\AMAS}{AMAS\xspace}
\newcommand{\Agents}{\mathcal{A}} 
\newcommand{\A}{\ensuremath{A}} 
\newcommand{\Loc}{\ensuremath{L}}
\renewcommand{\loc}{\ensuremath{l}}
\newcommand{\Evt}{\ensuremath{Evt}}
\newcommand{\evt}{\ensuremath{e}}
\newcommand{\Prot}{\ensuremath{R}}
\newcommand{\Trans}{\ensuremath{T}}
\newcommand{\PV}{\mathit{PV}}
\renewcommand{\Val}{\ensuremath{V}}

\renewcommand{\model}{\mathit{M}}
\renewcommand{\States}{\ensuremath{S}}

\newcommand{\seq}{\pi}
\newcommand{\strat}{\ensuremath{\sigma}}
\newcommand{\outcome}{\ensuremath{out}}

\newtheorem{definition}{Definition}
\newtheorem{remark}{Remark}

\title{Formal Verification of Probabilistic Multi-Agent Systems for Ballistic Rocket Flight Using Probabilistic Alternating-Time Temporal Logic}
\author[1,2]{Damian Kurpiewski}
\author[3]{Jędrzej Michalczyk}
\author[1,2]{Wojciech Jamroga}
\author[3]{\mbox{Jerzy Julian Michalski}}
\author[1]{Teofil Sidoruk}
\affil[1]{Institute of Computer Science, Polish Academy of Sciences}
\affil[2]{Nicolaus Copernicus University in Toruń, Poland}
\affil[3]{SpaceForest Sp. z o.o., Gdynia, Poland}

\date{}

\begin{document}

\maketitle

\begin{abstract}
This technical report presents a comprehensive formal verification approach for probabilistic agent systems modeling ballistic rocket flight trajectories using Probabilistic Alternating-Time Temporal Logic (PATL). We describe an innovative verification framework specifically designed for analyzing critical safety properties of ballistic rockets engineered to achieve microgravity conditions for scientific experimentation. Our model integrates authentic flight telemetry data encompassing velocity vectors, pitch angles, attitude parameters, and GPS coordinates to construct probabilistic state transition systems that rigorously account for environmental stochasticity, particularly meteorological variability. We formalize mission-critical safety properties through PATL specifications to systematically identify trajectory deviation states where the rocket risks landing in prohibited or hazardous zones. The verification framework facilitates real-time safety monitoring and enables automated intervention mechanisms, including emergency engine disengagement protocols, when predefined safety thresholds are exceeded. Experimental validation demonstrates the practical effectiveness and reliability of our approach in ensuring mission safety while maintaining scientific mission objectives.
\end{abstract}

\section{Introduction}
\label{sec:introduction}

Many important properties of agent systems refer to \emph{strategic abilities} of agents and their groups.
\emph{Alter\-nating-time temporal logic} \ATLs~\cite{Alur02ATL,Schobbens04ATL} and \emph{Strategy Logic} \SL~\cite{Mogavero14behavioral} provide powerful tools to reason about such aspects of MAS.
For example, the \ATLs formula $\coop{taxi}\Always\neg\prop{fatality}$ expresses that the autonomous cab can drive in such a way that no one gets ever killed. Similarly, $\coop{taxi,passg}\Sometm\prop{destination}$ says that the cab and the passenger have a joint strategy to arrive at the destination, no matter what the other agents do.
Specifications in agent logics can be used as input to algorithms and tools for \emph{model checking},
that have been in constant development for over 20 years~\cite{Alur98mocha-cav,Busard15reasoning,Cermak15mcmas-sl-one-goal,Huang14symbolic-epist,Kurpiewski21stv-demo,Lomuscio17mcmas}.

Model checking of strategic abilities is hard, both theoretically and in practice. First, it suffers from the well-known state/transition-space explosion.
Moreover, the space of possible strategies is at least exponential \emph{on top of the state-space explosion}, and incremental synthesis of strategies is not possible in general -- especially in the realistic case of agents with partial observability.
Even for the more restricted (and computation-friendly) logic \ATL, model checking of its imperfect information variants is \Deltwo- to \Pspace-complete for agents playing memoryless strategies~\cite{Bulling10verification,Schobbens04ATL} and \EXPTIME-complete to undecidable for agents with perfect recall~\cite{Dima11undecidable,Guelev11atl-distrknowldge}.
The theoretical results concur with outcomes of empirical studies on benchmarks~\cite{Busard15reasoning,Jamroga19fixpApprox-aij,Lomuscio17mcmas}, as well as recent attempts at verification of real-life multi-agent scenarios~\cite{Jamroga20Pret-Uppaal,Kurpiewski19embedded}.

This technical report presents a comprehensive case study demonstrating the application of state-of-the-art strategy synthesis techniques to the Perun Rocket system developed by SpaceForest. This work represents a significant methodological advancement as we bridge the gap between cutting-edge theoretical research in formal verification and real-world aerospace engineering challenges. The case study exemplifies the practical deployment of highly experimental and innovative formal methods (specifically, probabilistic model checking and strategic reasoning in agent systems) to a genuine aerospace scenario with critical safety requirements.

The verification of rocket flight systems poses unique challenges that distinguish it from traditional software verification tasks. Unlike conventional computational systems, ballistic rockets operate in highly dynamic, uncertain environments where atmospheric conditions, mechanical tolerances, and sensor inaccuracies introduce substantial stochasticity. Furthermore, the multi-agent perspective becomes crucial when considering the rocket's control systems, environmental factors, and ground-based monitoring as interacting entities whose collective behavior determines mission success or failure.

Our approach leverages PATL to express and verify complex strategic properties about what the rocket's control systems can achieve under various probabilistic scenarios. This enables us to reason formally about questions such as: ``Can the flight control system guarantee with probability at least $p$ that the rocket will remain within safe trajectory bounds?''.

The integration of formal methods with aerospace engineering represents not only a technical challenge but also an opportunity to establish new standards for safety verification in the emerging commercial space industry.

\section{What Agents Can Achieve}
\label{sec:logic}

In this section, we introduce the formalism of Asynchronous Multi-agent Systems (\AMAS) \cite{Jamroga20POR-JAIR},
as well as the syntax and semantics of Alternating-time Temporal Logic \ATLs \cite{Alur02ATL,Schobbens04ATL},
which allows for specifying relevant properties of SAI models,
in particular the \emph{strategic ability} of agents to enforce a goal.

\subsection{Asynchronous MAS}

\AMAS can be thought of as networks of automata, where each component corresponds to a single agent.

\begin{definition}[\AMAS \cite{Jamroga20POR-JAIR}]\label{def:amas}
An \emph{asynchronous multi-agent system} (\AMAS) consists of $n$ agents $\Agents = \set{1,\dots,n}$,
each associated with a 7-tuple $\A_i = (\Loc_i, \iota_i, \Evt_i, \Prot_i, \Trans_i, \PV_i, \Val_i)$, where:
\begin{itemize}
\item $\Loc_i=\{\loc_i^1,\ldots,\loc_i^{n_i}\} \neq \emptyset$ is a finite set of \emph{local states};
\item $\iota_i \in \Loc_i$ is an \emph{initial local state};
\item $\Evt_i=\{\evt_i^1,\ldots,\evt_i^{m_i}\} \neq \emptyset$ a finite set of \emph{events};
\item $\Prot_i: \Loc_i \to 2^{\Evt_i}\setminus\{\emptyset\}$ is a \emph{repertoire of choices},
			assigning available subsets of events to local states;
\item $\Trans_i: \Loc_i \times \Evt_i \fpart \Loc_i$ is a (partial) \emph{local transition function}
			such that $\Trans_i(\loc_i,\evt)$ is defined iff $\evt\in \Prot_i(\loc_i)$.
			That is, $\Trans_i(\loc,\evt)$ indicates the result of executing event $\evt$ in state $\loc$ from the perspective of agent $i$;
\item $\PV_i$ is a set of the agent's \emph{local propositions},
			with $\PV_j$, $\PV_k$ (for $j \neq k \in \Agents)$ assumed to be disjoint;
\item $\Val_i: \Loc_i \then \powerset{\PV_i}$ is a \emph{valuation function}.
\end{itemize}
Furthermore, we denote:
\begin{itemize}
\item by $\Evt = \bigcup_{i \in \Agents} \Evt_i$, the set of all events;
\item by $\Loc = \bigcup_{i \in \Agents} \Loc_i$, the set of all local states;
\item by $Agent(\evt) = \set{i \in \Agents \mid \evt \in \Evt_i}$, the set of all agents which have event $\evt$ in their repertoires;
\item by $\PV = \bigcup_{i \in \Agents} \PV_i$ the set of all local propositions.
\end{itemize}
\end{definition}

The \emph{model} of an \AMAS provides its execution semantics with asynchronous interleaving of private events and synchronisation on shared ones.

\begin{definition}[Model]\label{def:model}
The \emph{model} of an \AMAS is a 5-tuple $\model = (\Agents, \States, \iota, \Trans, \Val)$, where:
\begin{itemize}
\item $\Agents$ is the set of \emph{agents};
\item $\States \subseteq \Loc_1\times\ldots\times\Loc_n$ is the set of \emph{global states},
			including all states reachable from $\iota$ by $\Trans$ (see below);
\item $\iota = (\iota_1,\dots,\iota_n) \in \States$ is the \emph{initial global state};
\item $\Trans: \States\times\Evt \fpart \States$ is the \emph{global transition function},
			defined by $\Trans(\state_1,\evt) = \state_2$ iff $\Trans_i(\state_1^i,\evt) = \state^i_2$ for all $i \in Agent(\evt)$
			and $\state_1^i = \state^i_2$ for all $i \in \Agents \setminus Agent(\evt)$
			(where $\state_j^i \in \Loc_i$ is agent $i$'s local component of $\state_j$);
\item $\Val: \States \rightarrow 2^{\PV}$ is the \emph{global valuation function},
			defined as $\Val(\loc_1,\dots,\loc_n) = \bigcup_{i\in\Agents} \Val_i(\loc_i)$.
\end{itemize}
\end{definition}

\subsection{Strategic Ability}

Linear and branching-time temporal logics, such as \LTL and \CTLs~\cite{Emerson90temporal}, have long been used in formal verification.
They enable to express properties about \emph{how} the state of the system will (or should) evolve over time.
However, in systems that involve autonomous agents,
whether representing human users or AI components
it is usually of interest \emph{who} can direct its evolution a particular way.

\ATLs \cite{Alur02ATL} extends temporal logics with \emph{strategic modalities} that allow for reasoning about such properties.
The operator $\coop{\A}\gamma$ says that agents in group (coalition) $\A$ have a \emph{strategy} to enforce property $\gamma$.
That is, as long as agents in $\A$ select events according to the strategy, $\gamma$ will hold no matter what the other agents do.
\ATLs has been one of the most important and popular agent logics in the last two decades.

\begin{definition}[Syntax of \ATLs]\label{def:syntax}
The language of \ATLs is defined by the grammar:
\begin{center}
$\varphi::= \prop{p} \mid \neg \varphi \mid \varphi\wedge\varphi \mid \coop{\A}\gamma$, \\
$\gamma::=\varphi \mid \neg\gamma \mid \gamma\land\gamma \mid \Next\gamma \mid \gamma\Until\gamma$,
\end{center}
where $\prop{p} \in \PV$ and $\A \subseteq \Agents$.
The definitions of Boolean connectives and temporal operators $\Next$ (``next'') and $\Until$ (``strong until'') are standard;
remaining operators $\Release$ (``release''), $\Always$ (``always''), and $\Sometm$ (``sometime'') can be derived as usual.
\end{definition}

Various types of strategies can be defined, based on the state information and memory of past states available to agents \cite{Schobbens04ATL}.
In this work, we focus on \emph{imperfect information, imperfect recall strategies}.

\begin{definition}[Strategy]
A \emph{memoryless imperfect information strategy} for agent $i \in \Agents$ is a function
$\strat_i \colon \Loc_i \to 2^{\Evt_i}\setminus\emptyset$ such that $\strat_i(\loc) \in \Prot_i(\loc)$ for each local state $\loc \in \Loc_i$.
A \emph{joint strategy} $\strat_\A$ of coalition $\A \subseteq \Agents$ is a tuple of strategies $\strat_i$, one for each agent $i \in \A$.
\end{definition}

The outcome set of a strategy collects all paths consistent with it,
\ie that may occur when the coalition follows the chosen strategy,
while opposing agents choose freely from their protocols.

\begin{definition}[Outcome]\label{def:outcome}
The \emph{outcome} of strategy $\strat_\A$ in global state $\state \in \States$ of model $\model$,
denoted by $\outcome_\model(\state,\strat_\A)$,
is the set of all paths $\seq = \state_0 \evt_0 \state_1 \evt_1 \dots$, such that
$\seq \in \outcome_\model(\state,\strat_\A)$
iff $\state_0 = \state$, and for each $j \geq 0$
$$\begin{cases}
		\evt_j\in\strat_\A(\state_j^i) & \text{for every agent } i\in Agent(\evt_j)\cap \A, \\
		\evt_j\in\bigcup\Prot_i(\state_j^i) & \text{for every agent } i\in Agent(\evt_j)\setminus \A,
\end{cases}$$

\noindent
where $\state^i$ denotes the local component $\loc_i$ of $\state$.
\end{definition}

\begin{definition}[Asynchronous semantics of \ATLs \cite{Jamroga20POR-JAIR}]
The asynchronous semantics of the strategic modality in \ATLs is defined by the following clause:
\vspace{0.1cm}

\noindent
$\model,\state \satisf \coop{\A}\gamma$ iff there is a strategy $\strat_\A$ such that $\outcome_\model(\state,\strat_\A) \neq \emptyset$ and,
for each path $\seq\in \outcome_\model(\state,\strat_\A)$, we have $\model,\seq \satisf \gamma$.
\vspace{0.1cm}

\noindent
The remaining clauses for temporal operators and Boolean connectives are standard, see~\cite{Emerson90temporal}.
\end{definition}


\section{Strategic Ability in Stochastic MAS}\label{sec:patl}

We start by recalling the basic definitions of stochastic multi-agent models and strategic play~\cite{Boutilier99mmdp,Chen07PATL,Huang12probabilisticATL}.  In our presentation, we follow mainly~\cite{Belardinelli24PATL}.

\subsection{Probabilistic Models for MAS}\label{sec:probMAS}

Fix finite non-empty sets $\Agt$ of agents $a, a',\ldots$; $\Act$ of actions $\alpha, \alpha',\ldots$; and
$\Props$ of atomic propositions $p, p',\ldots$.
We  write $\profile{o}$ for a tuple $(o_{a})_{{a}\in\Agt}$ of objects, one for each agent; such tuples are called \emph{profiles}. A \emph{joint action} or \emph{move} $\mov$ is an element of $\Act^{\Agt}$.
Given a profile $\profile{o}$ and $\coalition\subseteq\Agt$, we let $o_\coalition$ be the components for the agents in  $\coalition$.
Moreover, we use $\Agt_{-\coalition}$ as a shorthand for $\Agt\setminus\coalition$. 

\smallskip
\para{Distributions.} Let $X$ be a finite non-empty set. A \emph{(probability) distribution} over $X$ is a function $\distribution:X \to [0,1]$ such that $\sum_{x \in X} \distribution(x) = 1$. $\Dist(X)$ is the set of distributions over $X$. We write $x \in \distribution$ for $\distribution(x) > 0$.
If $\distribution(x) = 1$ for some element $x \in X$, then $\distribution$ is a \emph{point (a.k.a. Dirac) distribution}.
 If $\distribution_i$ is a distribution over $X_i$, then, writing $X = \prod_{i} X_i$, the \emph{product distribution} of the $\distribution_i$ is the distribution $\distribution:X \to [0,1]$ defined by $\distribution(x) = \prod_{i} \distribution_i(x_i)$.

\para{Markov Chains. }
  A \emph{Markov chain} $M$ is a tuple $(\States,\distribution)$ where $\States$ is a set of states and $\distribution \in \Dist(\States \times \States)$ is a distribution. The values $\distribution(s,t)$ are called \emph{transition probabilities} of $M$.

\para{Stochastic Concurrent Game Structures. }
  A \emph{stochastic concurrent game structure with imperfect information} (or simply \emph{iCGS})
  $\System$ is a tuple 
  $(
  \States, \legal, 
  \trans, \val, \{\obsrel\}_{{a}\in\Agt})$ where
  (i) $\States$ is a finite, non-empty set of \emph{states};
   (ii) $\legal: \States \times \Agt \to 2^\Act\setminus\{\emptyset\}$ is a function defining the available actions for each agent in each state, i.e., the repertoires of choices. We write $\legal(\state)$ for the set of tuples $(\legal(\state, {a}))_{{a}\in\Agt}$. It is usually assumed that $\legal(\state,{a}) = \legal(\state',{a})$ whenever $\state \obsrel[a] \state'$ (see below);
  (iii)
  for each state $\state \in \States$ and each  move $\mov \in \legal(\state)$, the \emph{stochastic transition function} $\trans$ gives the (conditional) probability $\trans(\state, \mov)$ of a transition from state $\state$ for all $\state' \in \States$ if each player ${a} \in \Agt$ plays the action $\mov_a$; we also write this probability as $\trans(\state, \mov)(\state')$ to emphasize that $\trans(\state, \mov)$ is a probability distribution on $\States$;
  (iv) $\val:\States \to 2^{\Props}$ is a \emph{labelling function};
  (v)
    $\obsrel\;\subseteq \States\times\States$ is an equivalence relation called the  \emph{observation relation} of agent ${a}$.

A pointed \CGS is a pair $(\System,\state)$ where \mbox{$\state \in \States$} is a special state designed as initial.
Throughout this paper, we assume that iCGSs are {\em uniform}, that is, if two states are indistinguishable for an agent ${a}$, then ${a}$ has the same available actions in both states. Formally, if $\state \obsrel \state'$ then $\legal(\state, {a}) = \legal(\state', {a})$, for any $\state, \state' \in \States$ and ${a} \in \Agt$. For each state $\state \in \States$ and joint action $\mov \in \legal(\state)$, we also assume 
  that there is a state $\state'\in\States$ such that $\trans(\state, \mov)(\state')$ is non-zero, that is, every state has a successor state from a legal move.

Finally, we say that $\System$ is \emph{deterministic} (instead of stochastic) if every $\trans(\state,\mov)$ is a point distribution.

\para{Plays. }
A \emph{play} in a iCGS $\System$ is an infinite sequence $\path=\state_0 \state_1 \cdots$ of states
such that there exists a sequence $\mov_0 \mov_1 \cdots$ of joint-actions such that for every $i \geq 0$, $\mov_i \in \legal(\state_{i})$ and  $\state_{i+1} \in \trans(\state_i,\mov_i)$ (\ie, $\trans(\state_i,\mov_i)(\state_{i+1} )>0$).
We write $\path_i$ for state $\state_i$,
$\path_{\geq i}$ for the suffix of
$\path$ starting at position $i$.
Finite prefixes of plays are called \emph{histories}, and the set of all histories is denoted $\History$. Write $\last(\history)$ for the last state of a history $\history$.

\para{Strategies. }
A (general) \emph{probabilistic strategy} for agent ${a}\in\Agt$ is a  function $\str_{a}:\History \to  \Dist(\Act)$ that maps each history to a probability distribution over the agent's actions. It is required that $\str_{a}(\history)(\act) = 0$ if $\act \not \in \legal(\last(\history),{a})$.
We denote the set of ${a}$'s general strategies by $\setstrat_{a}$.

A \emph{memoryless uniform  probabilistic strategy} for an agent ${a}$ is a function $\sigma_{a}: \States \to \Dist(\Act)$, in which:
(i) for each $\state$, we have $\str_{a}(\state)(\act) = 0$ if $\act \not \in \legal(\state,{a})$; and
(ii) for all positions $\state,\state'$ such that $\state\obsrel\state'$, we have $\str_{a}(\state)=\str_{a}(\state')$. We let $\setstrat_a$ be the set of memoryless uniform strategies for agent ${a}$.
We call a memoryless strategy $\str_{a}$ \emph{deterministic} if $\str_{a}(\state)$ is a point distribution for every $\state$.

A \emph{collective strategy} for agents $A\subseteq\Agt$ is a tuple of strategies $\str_{a}$, one per agent ${a}\in A$.
We denote the set of $A$'s collective general strategies and memoryless uniform strategies, respectively, by $\setstrat_A$ and $\setstrat^r_A$.
Moreover, a \emph{strategy profile} is a tuple $\profile\str = \str_\Agt$ of strategies for all the agents. We write $\profile\str_{a}$ for the strategy of  ${a}$ in profile $\profile\str$.

\subsection{Probabilistic \ATL and \ATLs}
Now we present the syntax and semantics of the Probabilistic Alternating-time Temporal Logics \PATLs and \PATL~\cite{Chen07PATL,Huang12probabilisticATL,Belardinelli23PATL,Belardinelli24PATL}, interpreted under the assumption of imperfect information. Again, we follow~\cite{Belardinelli24PATL} in our presentation. Note that~\cite{Belardinelli24PATL} adopts the {\em objective} semantics of strategic ability, where the coalition is supposed to have a strategy that works from the initial state of the game. In contrast, \cite{Huang12probabilisticATL}~uses the {\em subjective} semantics of strategic ability, where the agents need a strategy that wins from the all the observationally equivalent states.\footnote{
  For a more thorough discussion of objective vs.~subjective ability, cf.~\cite{Bulling14comparing-jaamas,Agotnes15handbook}. }
In this paper, we consider both accounts, as they are equally relevant in the literature. In particular, we integrate the {\em objective} and {\em subjective} semantics of probabilistic ability into a single framework.

\begin{definition}[\PATLs]\label{def:ATLsF-syntax}
State formulas $\varphi$
and path formulas $\psi$
are defined by the following grammar, where $p \in \Props$, $\coalition \subseteq \Agt$, $d$ is a rational constant in
$[0, 1]$, and $\bowtie \in
\{\leq, <, >, \geq\}$:
\begin{eqnarray}
	\varphi  & ::= & p \mid \neg \varphi \mid  {\varphi \lor  \varphi} \mid \coop{\coalition}^{\bowtie d} \psi\\
 	\psi  & ::= & \varphi \mid \neg \psi \mid  {\psi \lor  \psi}  \mid \Next \psi \mid \psi \Until \psi \mid \psi \Release \psi
\end{eqnarray}
Formulas in \PATLs are all and only the state formulas $\varphi$.
\end{definition}

An important syntactic restriction of \PATLs, namely  \PATL, is obtained by restricting path formulas as follows:
\begin{eqnarray}
 	\psi  & ::= & \Next \varphi \mid \varphi \Until \varphi \mid \varphi \Release \varphi
\end{eqnarray}
which is tantamount to the following grammar for state formulas:
	\begin{align}   	
		\varphi ::= p \mid \neg \varphi  \mid  \varphi \lor \varphi \mid \coop{\coalition}^{\bowtie d} \Next \varphi \mid \coop{\coalition}^{\bowtie d}(\varphi \Until \varphi)
  \mid \coop{\coalition}^{\bowtie d}(\varphi \Release \varphi)
  \end{align}
where again $p \in \Props$, $\coalition \subseteq \Agt$, and $\bowtie \in \{\leq, <, >, \geq\}$.

Formulas of \PATL and \PATLs are interpreted over iCGSs. 

\para{Probability Space on Outcomes. }
An \emph{outcome} of a strategy $\str_A$ and a state $\state$
is a set of probability distributions over infinite paths, defined as follows.

First, by an \emph{outcome path of a strategy profile $\profile\str$ and state $\state$}, we refer to every play $\path$ that starts with $\state$ and is extended by letting each agent follow their strategies in $\profile\str$, i.e.,
$\path_{0} = \state$, and for every $k \geq 0$ there exists $\mov_k \in
\profile\str(\path_k)$ such that $\path_{k+1} \in \trans(\path_k,\mov_k)$.
The set of {outcome paths} of strategy profile $\profile\str$ and state $\state$ is denoted as $outpaths(\profile\str,\state)$.
A given iCGS $\System$, strategy profile $\profile{\str}$, and state $\state$ induce an infinite-state Markov chain
$M_{\profile{\str},\state}$ whose states are the finite prefixes of plays in $outpaths(\profile{\str},\state)$.
Such finite prefixes of plays are actually \emph{histories}.
Transition probabilities in  $M_{\profile{\str},\state}$ are defined as  $p(\history,\history\state')=\sum_{\mov\in\Act^\Agt}
\profile{\str}(\history)(\mov) \cdot \trans(\last(\history),\mov)(\state')$.
The Markov chain $M_{\profile{\str},\state}$ induces a canonical probability space on
its set of infinite paths~\cite{Kemeny76stochastic}, and thus also on $outpaths(\profile{\str},\state)$.~\footnote{
  This is a classical construction, see for instance~\cite{Clarke18principles,Berthon20alternating}. } 

Given a coalitional strategy $\str_{\coalition} \in \prod_{{a} \in \coalition} \setstrat_a$, we
define its {\em objective outcome} from state $\state \in \States$ as the set
$out_{o,\coalition}(\str_{\coalition},\state) =
  \{out((\str_{\coalition},\str_{\Agt\setminus\coalition}),\state) \mid \str_{\Agt\setminus\coalition} \in \setstrat_{\Agt\setminus\coalition} \}$
of probability measures consistent with strategy $\str_{\coalition}$ of the players in $\coalition$.
Note that the opponents can use any general strategy for $\str_{\Agt\setminus\coalition}$, even if $\coalition$ must employ only uniform memoryless strategies for $\str_{\coalition}$.

The {\em subjective outcomes} are then defined as the set
\begin{equation}
out_{s,\coalition}(\str_{\coalition},\state) = \bigcup_{\state' \sim_a \state, a \in \coalition} out_{o, \coalition}(\str_\coalition,\state') \label{subj}
\end{equation}

We will use $\mu^{\str_{\coalition}}_{x,\state}$ to range over the elements of $out_{x,\coalition}(\str_{\coalition},\state)$, for $x \in \{s, o\}$.

\para{Semantics.}
For $x$ equal to either $s$ or $o$, state and path formulas in \PATLs are interpreted in a iCGS $\System$
  and a state $\state$, resp.~path $\path$, according to the $x$-interpretation of strategy operators, as follows (clauses for Boolean connectives are omitted as immediate):
\begingroup
\allowdisplaybreaks
\begin{align*}
 \System,\state &\models_x p & \text{ iff } & p \in \val(\state)\\
\System, \state  & \models_x \coop{\coalition}^{\bowtie d} \psi & \text{ iff }  & 
\exists \profile{\str_{\!\coalition}} \in \prod_{{a} \in\coalition} \setstrat_a  \text{ such that }
\\ & & & \forall \mu^{\profile{\str_{\!\coalition}}}_{x,\state} \in out_{x,\coalition}(\profile{\str_{\!\coalition}},\state)
\text{, }
\\ & & &
\mu^{\profile{\str_{\!\coalition}}}_{x,s}(\{\path \mid \System,\path \models_x \psi\}) \bowtie d \\
 \System,\path &\models_x \Next \psi & \text{ iff } & \System,\path_{\geq 1} \models_x \psi
 \\
\System, \path  & \models_x \psi_1 \Until \psi_2 & \text{ iff } &  \exists k \geq 0 \text{ s.t. } \System,\path_{\geq k} \models_x \psi_2 \text{ and }
\\ & & &
\forall j \in [0,k)\, \,  \System,\path_{\geq j}\models_x \psi_1\\
\System, \path  & \models_x \psi_1 \Release \psi_2 & \text{ iff } &  \forall k \geq 0, \System,\path_{\geq k} \models_x \psi_2 \text{ or }
\\ & & &
\exists j \in [0,k) \text{ s.t. }  \System,\path_{\geq j}\models_x \psi_1
\end{align*}
\endgroup

\begin{remark}
Note that standard \ATLs formulas $\coop{\coalition}\psi$ can be interpreted in stochastic iCGS by simply ignoring the probabilities of transitions, i.e., via a projection of the probabilistic transition relation $\trans$ to the non-probabilistic relation $T$ defined by $(\state,\mov,\state') \in T$ iff $\trans(\state, \mov)(\state') > 0$.
\end{remark}

\begin{remark}
Moreover, the logic of Probabilistic \CTLs (\PCTLs) can be embedded in \PATLs by assuming $\Probpath^{\bowtie d} \psi \equiv \coop{\emptyset}^{\bowtie d} \psi$.
\end{remark}


\section{Case Study: Perun Ballistic Rocket by SpaceForest}\label{sec:system-model}

\noindent This section provides an overview of the Perun ballistic rocket developed by SpaceForest, a Polish technology company specializing in rocket technologies and space systems. We describe the company's background, the technical specifications of the Perun rocket, its payload section designed for scientific experiments, and the research applications it supports.

\subsection{Company Background}
SpaceForest is a Polish technology company headquartered in Gdynia, specializing in rocket technologies, radars and SAR (Synthetic Aperture Radar), advanced electronics, microwave filters, and composites. Founded in 2004, the company has accumulated 21 years of experience in space technologies, with over 50 dedicated employees (95\% of the team consists of technically qualified personnel). Its business profile spans the space, aviation, defense industries, and advanced electronic and material technologies.

\subsection{Technical Specifications}
\begin{table}[h]
\centering
\caption{Basic technical parameters of the PERUN rocket}
\begin{tabular}{ll}
\toprule
Parameter & Value \\
\midrule
Height & 11.5m \\
Diameter & 0.45m \\
Launch mass & 970kg \\
Apogee & 150km \\
Time in space & approx. 5min \\
Payload capacity & 50kg \\
Internal payload bay diameter & 444mm \\
Internal payload bay length & 830mm \\
Propulsion type & Eco-friendly rocket engine \\
\bottomrule
\end{tabular}
\label{tab:perun_specs}
\end{table}

\noindent The PERUN rocket is a two-stage, hybrid engine (solid fuel and liquid oxidizer) suborbital vehicle designed to carry scientific and technological payloads to altitudes of up to 150km. It features a recoverable payload section, advanced telemetry and control systems, and a low acceleration profile suitable for sensitive experiments. Technical specifications of the PERUN rocket are summarized in Table \ref{tab:perun_specs}.

\paragraph{Key Systems and Technologies}

\noindent The PERUN rocket incorporates several advanced systems and technologies to ensure mission success and safety:

\begin{itemize}
    \item \textbf{Telecommunication system:} Two-way real-time data transmission with in-flight mission parameter modification capability
    \item \textbf{Thrust vector control (VTC) system:} Precision engine thrust management for optimal flight trajectory
    \item \textbf{Telemetry and control:} Comprehensive flight parameter monitoring and remote command system
    \item \textbf{Recoverable payload section:} Engineered for safe payload return to Earth
    \item \textbf{Low acceleration profile:} Optimized launch conditions for sensitive experimental payloads
\end{itemize}

\subsection{Payload Section and Experiment Integration}

\noindent The PERUN rocket's payload section is specifically designed to support a variety of scientific and technological experiments in microgravity conditions.

\paragraph{Payload Section Parameters}
\begin{itemize}
    \item Nominal payload mass: 50kg
    \item Research altitude: 150km
    \item Microgravity duration: approx. 5min
    \item Maximum payload dimensions: 830mm length $\times$ 444mm diameter
    \item Payload types supported: Active or passive experimental configurations
\end{itemize}

\paragraph{Integration Process}
\begin{itemize}
    \item Mission-specific payload adaptation and qualification
    \item Late access capability to payload section prior to launch
    \item Clean room handling environment (ISO 7 class)
    \item Flexible launch options from designated terrestrial locations
\end{itemize}

\subsection{Research Applications}

\noindent The PERUN rocket's payload section is designed to accommodate a wide range of scientific and technological experiments, enabling researchers to explore various phenomena in microgravity.

\paragraph{Scientific Research Domains}
\begin{itemize}
    \item \textbf{Life sciences and medicine:} Cellular and tissue behavior in microgravity; pharmaceutical validation
    \item \textbf{Human physiology:} Neural and immune cell response to weightlessness simulation
    \item \textbf{Chemistry:} Convection-free reaction analysis; crystallization kinetics studies
    \item \textbf{Physics:} Fundamental theory verification in weightlessness and high-acceleration environments
    \item \textbf{Fluid dynamics and combustion:} Liquid/gas behavior and fire propagation in microgravity
    \item \textbf{Materials science:} Crystallization processes and novel material properties in zero-gravity
\end{itemize}

\paragraph{Technological Validation Applications}
\begin{itemize}
    \item Pre-orbital validation of electronic components, sensors, and mechanical systems
    \item Space construction techniques: Assembly methods and robotics testing in weightlessness
    \item Upper atmospheric research: Data collection from regions inaccessible to balloons and drones
    \item Astronomical instrumentation: Testing optical systems above atmospheric interference
\end{itemize}

\subsection{Mission Timeline and Development}

\noindent The PERUN rocket is currently in the advanced testing phase, with a series of planned test flights and a roadmap for operational deployment.

\paragraph{Test Flight Schedule (2025-2026)}
\begin{itemize}
    \item October 2025: Ustka test flight (90\% payload capacity utilization)
    \item November 2025: Ustka test flight (75\% payload capacity utilization)
    \item First half 2026: Portuguese test flight (50\% payload capacity utilization)
    \item Second half 2026: Danish sea platform test flight (45\% payload capacity utilization)
\end{itemize}

\paragraph{Operational Development Path}
\begin{itemize}
    \item 2026-2027: Commercialization phase with payload reservations and exclusive mission offerings
    \item 2026-2027: Implementation of land recovery systems
    \item 2026-2027: Pursuit of Polish spaceport access or foreign launch site agreements
    \item Post-2027: Development of enhanced PERUN variant for 300km altitude operations
\end{itemize}

\subsection{Representative Research Missions}

\noindent Below are selected experiments planned for the PERUN rocket's payload section, showcasing its scientific and technological capabilities.

\paragraph{MXene in LEO (AGH University of Science and Technology)}
Testing six MXene-based sensors to evaluate material response to suborbital flight conditions (thermal variations, vibrations, pressure changes) under Prof. Tadeusz Uhl's leadership.

\paragraph{THOR (Analog Astronaut Training Center)}
Investigation of acceleration effects on biological growth media under Dr. Agata Kołodziejczyk, utilizing PERUN's late-access capability for biotechnology applications.

\paragraph{AstroFarms}
Microgravity analysis of symbiotic bacterial-yeast cultures (SCOBY) examining survivability, cellulose production, metabolic changes, and potential space applications of kombucha fermentation.

\paragraph{Lab-on-a-Chip}
World-first experiment with living bone cells on microfluidic platform assessing technology resilience to launch stresses and cellular survival in flight conditions.

\paragraph{Toraf}
Germination analysis of seeds exposed to suborbital conditions (extreme cold, radiation, low pressure) compared with terrestrial control samples.


\section{6-DOF Rocket Flight Simulator: Technical Description and Capabilities}\label{sec:simulator}

The 6-DOF (Six Degrees of Freedom) Rocket Flight Simulator is an in-house developed simulation environment created by SpaceForest for high-fidelity modeling and analysis of sounding rocket trajectories, dynamics, and control systems. This comprehensive simulation platform has been validated against actual SpaceForest flight data, providing reliable prediction capabilities throughout the entire mission profile from launch through recovery.

\subsection{Simulation Architecture}

The simulator is built using a modular architecture that allows for flexible configuration of vehicle parameters, environmental conditions, and mission profiles. It integrates various physical models, numerical solvers, and data processing tools to accurately replicate the complex interactions affecting rocket flight.

\begin{itemize}
    \item \textbf{Six Degrees of Freedom Model:} Complete dynamic model resolving translational motion (position, velocity, acceleration) and rotational motion (attitude, angular rates, angular acceleration)
    \item \textbf{Coordinate Systems:} Multiple reference frames including inertial, body-fixed, wind reference, and geodetic coordinate systems
    \item \textbf{Physical Models:}
    \begin{itemize}
        \item \textit{Propulsion System:} Thrust magnitude and direction, propellant mass depletion, Thrust Vector Control (TVC) dynamics, motor performance characteristics
        \item \textit{Aerodynamics:} Force and moment coefficients, Mach number and angle of attack dependencies, center of pressure modeling, atmospheric density variation
        \item \textit{Environmental Models:} Atmospheric conditions (temperature, pressure, density), gravitational field modeling, wind models (steady, gusts, turbulence)
    \end{itemize}
\end{itemize}

\subsection{Mission Event Modeling}

The simulator includes detailed event modeling capabilities to replicate critical mission phases and transitions:

\begin{itemize}
    \item \textbf{Stage Separation:}
    \begin{itemize}
        \item Payload-booster separation mechanics
        \item Separation impulse modeling
        \item Post-separation trajectory divergence
        \item Independent tracking of separated components
    \end{itemize}
    \item \textbf{Recovery System:}
    \begin{itemize}
        \item Multi-stage parachute deployment
        \item Drogue and main parachute sequencing
        \item Drag coefficient modeling
        \item Descent rate calculations
        \item Touchdown condition prediction
    \end{itemize}
\end{itemize}

\subsection{Sensor Simulation Capabilities}

The simulator incorporates realistic sensor models to generate synthetic measurements for onboard navigation and control systems:

\begin{itemize}
    \item \textbf{IMU Simulation:}
    \begin{itemize}
        \item 3-axis accelerometer and gyroscope modeling
        \item Measurement noise (white noise, random walk)
        \item Bias instability and scale factor errors
        \item Sensor misalignment errors
        \item Saturation limits
    \end{itemize}
    \item \textbf{GPS Receiver Simulation:}
    \begin{itemize}
        \item Realistic accuracy characteristics
        \item Variable update rate modeling
        \item Signal obstruction scenarios
    \end{itemize}
\end{itemize}

\subsection{Control System Development Environment}

The simulator provides a robust environment for developing and testing rocket guidance, navigation, and control (GNC) algorithms:

\begin{itemize}
    \item Trajectory design and optimization tools
    \item Guidance algorithm implementation framework
    \item Attitude control system design capabilities
    \item Thrust vector control algorithm development
    \item Stability augmentation system modeling
    \item Control law design and gain tuning interfaces
    \item Stability analysis across the complete flight envelope
\end{itemize}

\subsection{Software-in-the-Loop (SIL) Capabilities}

The simulator supports Software-in-the-Loop testing, enabling integration of actual flight software with the simulation environment:

\begin{itemize}
    \item Real-time or faster-than-real-time simulation execution
    \item Complete closed-loop vehicle and control integration
    \item Flight software reads simulated sensor data as if from actual hardware
    \item Control algorithms compute actuator commands based on simulated conditions
    \item Iterative execution through entire mission profile
    \item Automated test case execution framework
    \item Monte Carlo simulation for statistical analysis
    \item Fault injection and failure mode testing scenarios
    \item Regression testing capabilities for software updates
\end{itemize}

\subsection{Validation and Verification}

The simulator has undergone extensive validation against actual SpaceForest flight data to ensure accuracy and reliability:

\begin{itemize}
    \item Extensive validation against SpaceForest flight data
    \item Trajectory reconstruction from telemetry comparisons
    \item Model parameter tuning based on flight measurements
    \item Position, velocity, and attitude accuracy validation
    \item Event timing correlation (separation, parachute deployment)
    \item Descent rate and landing location prediction verification
    \item Continuous improvement through iterative flight data integration
\end{itemize}

\subsection{Primary Applications}

The 6-DOF Rocket Flight Simulator serves as a critical tool throughout the rocket development lifecycle:

\begin{itemize}
    \item Mission design and trajectory optimization
    \item Control system development and validation
    \item Flight software verification through SIL testing
    \item Monte Carlo analysis for comprehensive risk assessment
    \item Sensitivity studies and parametric analysis
    \item Post-flight analysis and correlation with mission data
    \item Pre-flight mission rehearsal and verification
    \item Crew training and procedure validation
\end{itemize}

\subsection{Operational Implementation}

The simulator has been successfully integrated into SpaceForest's development workflow, supporting multiple rocket development programs including the PERUN suborbital vehicle. Its modular architecture allows for configuration specific to different vehicle types while maintaining core simulation capabilities. The system supports both interactive simulation sessions for design exploration and automated batch processing for comprehensive parameter studies.

\begin{figure}[h]
\centering
\includegraphics[width=0.8\textwidth]{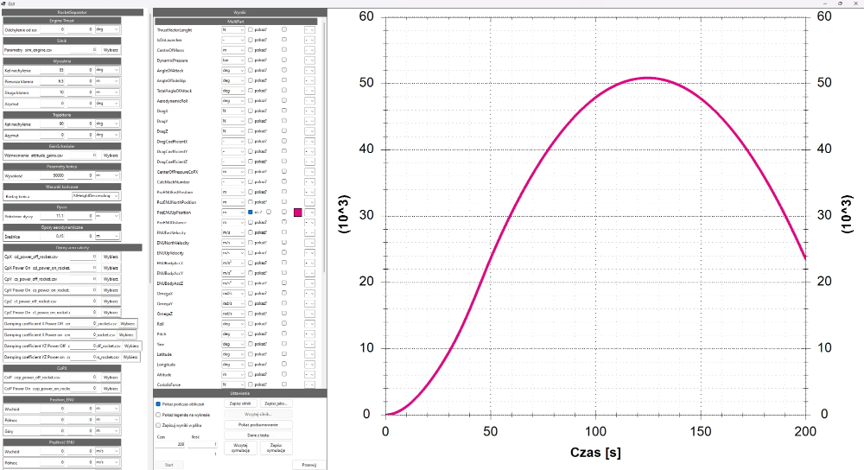}
\caption{High-fidelity simulator architecture and sample output used to generate synthetic rocket trajectories.}
\label{fig:simulator}
\end{figure}

\noindent Figure~\ref{fig:simulator} illustrates the simulator components and example outputs referenced in the simulator and data sources description.


\section{Formal Modelling of a Rocket}\label{sec:modeling}

\noindent Our verification framework models the ballistic rocket system as a \textbf{probabilistic state transition system} $\mathcal{M} = (S, s_0, Act, P, L)$ where:
\begin{itemize}
    \item $S$ is the set of states representing rocket configurations at discrete time steps
    \item $s_0 \in S$ is the initial launch state
    \item $Act = \{\text{continue}, \text{disengage}\}$ is the action set (engine continuation or disengagement)
    \item $P: S \times Act \times S \to [0,1]$ defines probabilistic state transitions
    \item $L: S \to 2^{AP}$ labels states with atomic propositions
\end{itemize}

\subsection{State Representation}
Each state $s \in S$ encapsulates the rocket's physical configuration at time $t$ (seconds since launch) as a tuple:
\[
s = \langle t,\ v,\ p,\ \theta,\ \phi,\ \psi \rangle
\]
where:
\begin{itemize}
    \item $v \in \mathbb{R}^3$ is the 3D velocity vector (m/s) from inertial measurement units
    \item $p = (lat,\ lon,\ alt) \in \mathbb{R}^3$ is the GPS position
    \item $\theta$ = pitch angle (rad), $\phi$ = roll angle (rad), $\psi$ = yaw angle (rad) from attitude sensors
\end{itemize}

We compare each observed state to a corresponding reference (ideal) state at the same time using a single deviation score that combines position, velocity, and attitude differences. States are classified by thresholding this deviation: a GoodState means the deviation is below a design-specified "good" threshold, and a BadState means the deviation exceeds a larger, design-specified "bad" threshold. Threshold values are chosen as design parameters (for example, the good threshold may correspond to a position-equivalent deviation of about 500m). These classifications depend only on the instantaneous, time-aligned deviation and do not rely on landing-probability estimates.

\subsection{Data Acquisition and Model Construction}
The transition probabilities $P$ are derived from two complementary data sources:
\begin{enumerate}
    \item \textbf{Historical test flights}: launches providing real-world trajectory data under documented weather conditions
    \item \textbf{High-fidelity simulator}: Company-developed computational simulation model that generates synthetic trajectories under randomized atmospheric conditions:
    \begin{itemize}
        \item Weather parameters sampled from historical distributions
        \item simulated trajectories covering edge cases beyond test flight capabilities
        \item Simulator validated against test flight data
    \end{itemize}
\end{enumerate}
All sensor data is time-synchronized and discretized into $k$-second intervals to create state sequences $\{s_0, s_1, \dots, s_T\}$, where $k$ is a design parameter of the model.

\subsection{Probabilistic Transition Modeling}
Transition probabilities are empirically derived through \textbf{frequentist counting} over the combined dataset:
\[
P(s, a, s') = \frac{N(s, a, s')}{\sum_{s'' \in S} N(s, a, s'')}
\]
where $N(s, a, s')$ counts occurrences of transition $s \xrightarrow{a} s'$ across all trajectories.

\section{Results and Analysis}\label{sec:results}

We have created three distinct models of the rocket system:
\begin{itemize}
    \item \textbf{Model A}: perfect information model without probabilistic transitions (nondeterministic)
    \item \textbf{Model B}: perfect information model with probabilistic transitions derived from the simulator data
    \item \textbf{Model C}: imperfect information model with probabilistic transitions, where the rocket has only partial observability of its state (e.g., noisy sensor readings), and can read only integer approximations from sensors, introducing uncertainty in state estimation
\end{itemize}

\begin{figure}[!t]
\centering
\includegraphics[scale=0.12]{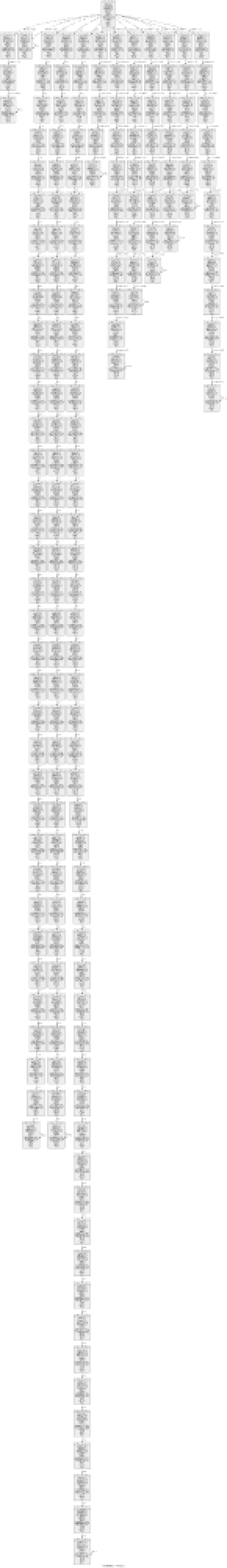}
\caption{Model A state transition diagram for time precision 6.0s (simplified for illustration)}
\label{fig:rocket_modelA_6s}
\end{figure}

\begin{figure}[t]
\centering
\includegraphics[scale=0.3]{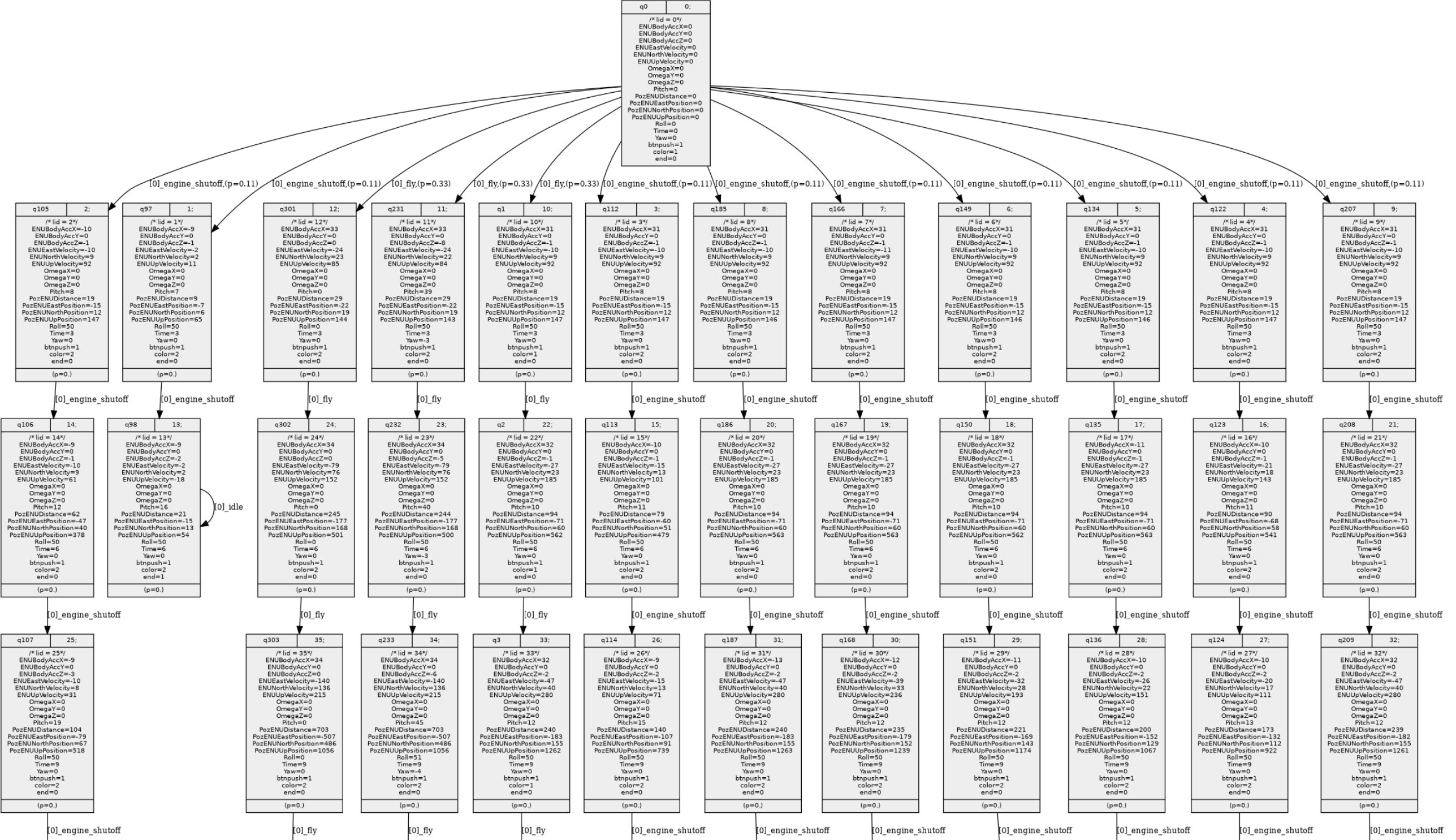}
\caption{Top part of the Model A state transition diagram for time precision 3.0s (simplified for illustration)}
\label{fig:rocket_top_modelA_3s}
\end{figure}

\begin{figure}[t]
\centering
\includegraphics[scale=0.3]{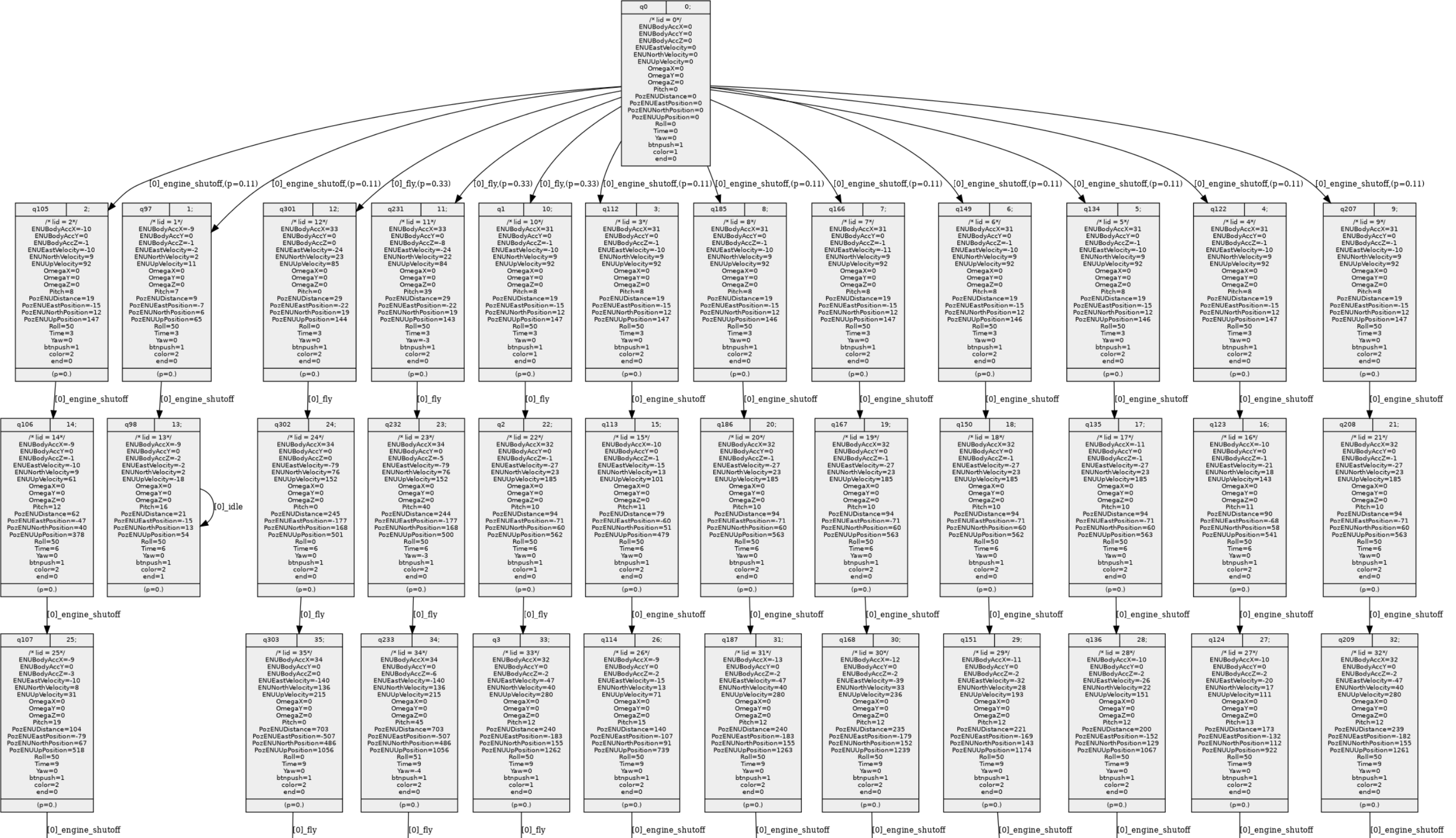}
\caption{Top part of the Model B state transition diagram for time precision 3.0s (simplified for illustration)}
\label{fig:rocket_top_modelB_3s}
\end{figure}

Classes of models are defined according to the time precision parameter. Since data from the simulator and test flights is recorded at 0.001s intervals, we aggregate states into coarser time steps of 1.0s, 0.5s, 0.2s, and 0.1s to create models of varying granularity. Finer time precision leads to larger state spaces and more detailed transition dynamics.

Top portions of the state transition diagrams for Models A and B at time precision of 3.0s are shown in Figures~\ref{fig:rocket_top_modelA_3s} and~\ref{fig:rocket_top_modelB_3s}, respectively. These diagrams illustrate the branching structure of possible rocket states and transitions over time.

Example state transition diagrams for Model A at time precision of 6.0s is shown in Figure~\ref{fig:rocket_modelA_6s}. As the time precision increases, the number of states and transitions grows significantly, capturing more nuanced rocket behavior.

We specify critical safety properties using PATL formulas to identify states where the rocket deviates from the desired trajectory and risks landing in restricted areas. We define several key properties. We verified these properties on all three models using the STV model checker. For Model A we used the standard ATL model checking approach, while for Models B and C we employed the probabilistic extension of ATL (PATL) to account for stochastic transitions. In all cases we set three probability thresholds: $p=0.9$ for high-confidence properties, $p=0.5$ for moderate-confidence properties and $p=0.25$ for low-confidence properties.

In all of our experiments the platform for model checking was a machine with an AMD Ryzen 7 5700X3D CPU @ 3.0GHz and 64GB of RAM. All times are given in seconds. Timeout was set to 4 hours.

\subsection{Property 1: Eventual Good State and Finish}

For model A we verify that the rocket can eventually reach a good state and complete the mission:

$\varphi_{1} = \langle\langle \text{Rocket} \rangle\rangle \mathbf{F} (\text{GoodState} \land \text{Finish})$

For models B and C we verify that the rocket can eventually reach a good state and complete the mission with at least probability $p$:

$\varphi_{1p} = \langle\langle \text{Rocket} \rangle\rangle^{\geq p} \mathbf{F} (\text{GoodState} \land \text{Finish})$

\paragraph{Model A Results}

\begin{table}[h]
\begin{tabular}{|c|c|c|c|c|}
\hline
\textbf{time precision} & \textbf{\#states} & \textbf{model gen. time} & \textbf{verif. time} & \textbf{verif. result} \\ \hline
1.0                     & 1102              & 0.102                    & 0.008                & TRUE                   \\ \hline
0.5                     & 2209              & 0.213                    & 0.023                & TRUE                   \\ \hline
0.2                     & 5602              & 0.801                    & 0.089                & TRUE                   \\ \hline
0.1                     & 11210             & 2.370                    & 0.119                & TRUE                   \\ \hline
\end{tabular}
\caption{Model A verification results for Property 1 at varying time precisions.}
\label{tab:modelA_prop1}
\end{table}

Results for Model A are summarized in Table~\ref{tab:modelA_prop1}. The model generation time, verification time, and verification result (TRUE/FALSE) are reported for each time precision. As we can observe the size of the model increases linearly with finer time precision, but verification remains efficient. The formula was verified as TRUE in all cases.

\paragraph{Model B Results}

\begin{table}[]
\begin{tabular}{|c|c|c|cc|cc|cc|}
\hline
\multirow{2}{*}{\textbf{\begin{tabular}[c]{@{}c@{}}time \\ precision\end{tabular}}} & \multirow{2}{*}{\textbf{\#states}} & \multirow{2}{*}{\textbf{\begin{tabular}[c]{@{}c@{}}m. gen. \\ time\end{tabular}}} & \multicolumn{2}{c|}{\textbf{p=0.25}}                       & \multicolumn{2}{c|}{\textbf{p=0.5}}                        & \multicolumn{2}{c|}{\textbf{p=0.9}}                        \\ \cline{4-9} 
                                                                                    &                                    &                                        & \multicolumn{1}{c|}{\textbf{v. time}} & \textbf{v. result} & \multicolumn{1}{c|}{\textbf{v. time}} & \textbf{v. result} & \multicolumn{1}{c|}{\textbf{v. time}} & \textbf{v. result} \\ \hline
1.0                                                                                 & 1102                               & 0.099                                  & \multicolumn{1}{c|}{4.482}            & TRUE                 & \multicolumn{1}{c|}{4.636}            & TRUE               & \multicolumn{1}{c|}{6.183}            & FALSE              \\ \hline
0.5                                                                                 & 2209                               & 0.216                                  & \multicolumn{1}{c|}{59.652}            & TRUE                & \multicolumn{1}{c|}{76.702}            & FALSE              & \multicolumn{1}{c|}{76.659}            & FALSE              \\ \hline
0.2                                                                                 & 5602                               & 0.801                                  & \multicolumn{1}{c|}{1022.352}           &  TRUE               & \multicolumn{1}{c|}{1350.994}           &  FALSE               & \multicolumn{1}{c|}{1643.021}          &  FALSE               \\ \hline
0.1                                                                                 & 11210                              & 2.388                                  & \multicolumn{1}{c|}{timeout}                 &      -              & \multicolumn{1}{c|}{timeout}                 &       -             & \multicolumn{1}{c|}{timeout}                 &       -             \\ \hline
\end{tabular}
\caption{Model B verification results for Property 1 at varying time precisions.}
\label{tab:modelB_prop1}
\end{table}

Results for Model B are summarized in Table~\ref{tab:modelB_prop1}. For each time precision, we report the number of states, model generation time, verification time, and verification result for probability thresholds $p=0.25$, $p=0.5$, and $p=0.9$. As we can see, verification times increase significantly with finer time precision, and the property is verified as TRUE at $p=0.25$ and also at $p=0.5$ but only for the coarsest model. For higher precision models we reach timeout for the verification.

\paragraph{Model C Results}

\begin{table}[]
\begin{tabular}{|c|c|c|cc|cc|cc|}
\hline
\multirow{2}{*}{\textbf{\begin{tabular}[c]{@{}c@{}}time \\ precision\end{tabular}}} & \multirow{2}{*}{\textbf{\#states}} & \multirow{2}{*}{\textbf{\begin{tabular}[c]{@{}c@{}}m. gen. \\ time\end{tabular}}} & \multicolumn{2}{c|}{\textbf{p=0.25}}                       & \multicolumn{2}{c|}{\textbf{p=0.5}}                        & \multicolumn{2}{c|}{\textbf{p=0.9}}                        \\ \cline{4-9} 
                                                                                    &                                    &                                        & \multicolumn{1}{c|}{\textbf{v. time}} & \textbf{v. result} & \multicolumn{1}{c|}{\textbf{v. time}} & \textbf{v. result} & \multicolumn{1}{c|}{\textbf{v. time}} & \textbf{v. result} \\ \hline
1.0                                                                                 & 1102                               & 0.099                                  & \multicolumn{1}{c|}{6.470}            & TRUE                 & \multicolumn{1}{c|}{6.861}            & FALSE               & \multicolumn{1}{c|}{9.377}            & FALSE              \\ \hline
0.5                                                                                 & 2209                               & 0.216                                  & \multicolumn{1}{c|}{80.493}            & TRUE                & \multicolumn{1}{c|}{95.028}            & FALSE              & \multicolumn{1}{c|}{96.904}            & FALSE              \\ \hline
0.2                                                                                 & 5602                               & 0.801                                  & \multicolumn{1}{c|}{1801.057}           &  TRUE               & \multicolumn{1}{c|}{2061.072}           &  FALSE               & \multicolumn{1}{c|}{2106.082}          &  FALSE               \\ \hline
0.1                                                                                 & 11210                              & 2.388                                  & \multicolumn{1}{c|}{timeout}                 &      -              & \multicolumn{1}{c|}{timeout}                 &       -             & \multicolumn{1}{c|}{timeout}                 &       -             \\ \hline
\end{tabular}
\caption{Model C verification results for Property 1 at varying time precisions.}
\label{tab:modelC_prop1}
\end{table}

Results for Model C are summarized in Table~\ref{tab:modelC_prop1}. For each time precision, we report the number of states, model generation time, verification time, and verification result for probability thresholds $p=0.25$, $p=0.5$, and $p=0.9$. As we can see, verification times increase significantly with finer time precision, and the property is verified as TRUE only at $p=0.25$. Similarly to Model B, for higher precision models we reach timeout for the verification.

\subsection{Property 2: Disengagement Safety}

For model A we verify that if the rocket disengages, it will remain in a good state:

$\varphi_2 = \langle\langle \text{Rocket} \rangle\rangle \mathbf{G} (\text{Disengaged} \implies \text{GoodState})$

For models B and C we verify that if the rocket disengages, it will remain in a good state with at least probability $p$:

$\varphi_{2p} = \langle\langle \text{Rocket} \rangle\rangle^{\geq p} \mathbf{G} (\text{Disengaged} \implies \text{GoodState})$

\paragraph{Model A Results}

\begin{table}[h]
\begin{tabular}{|c|c|c|c|c|}
\hline
\textbf{time precision} & \textbf{\#states} & \textbf{model gen. time} & \textbf{verif. time} & \textbf{verif. result} \\ \hline
1.0                     & 1102              & 0.102                    & 0.007                & TRUE                   \\ \hline
0.5                     & 2209              & 0.213                    & 0.011                & TRUE                   \\ \hline
0.2                     & 5602              & 0.801                    & 0.028                & TRUE                   \\ \hline
0.1                     & 11210             & 2.370                    & 0.077                & TRUE                   \\ \hline
\end{tabular}
\caption{Model A verification results for Property 2 at varying time precisions.}
\label{tab:modelA_prop2}
\end{table}

Results for Model A are summarized in Table~\ref{tab:modelA_prop2}. The model generation time, verification time, and verification result (TRUE/FALSE) are reported for each time precision. The formula was verified as TRUE in all cases.

\paragraph{Model B Results}

\begin{table}[]
\begin{tabular}{|c|c|c|cc|cc|cc|}
\hline
\multirow{2}{*}{\textbf{\begin{tabular}[c]{@{}c@{}}time \\ precision\end{tabular}}} & \multirow{2}{*}{\textbf{\#states}} & \multirow{2}{*}{\textbf{\begin{tabular}[c]{@{}c@{}}m. gen. \\ time\end{tabular}}} & \multicolumn{2}{c|}{\textbf{p=0.25}}                       & \multicolumn{2}{c|}{\textbf{p=0.5}}                        & \multicolumn{2}{c|}{\textbf{p=0.9}}                        \\ \cline{4-9} 
                                                                                    &                                    &                                        & \multicolumn{1}{c|}{\textbf{v. time}} & \textbf{v. result} & \multicolumn{1}{c|}{\textbf{v. time}} & \textbf{v. result} & \multicolumn{1}{c|}{\textbf{v. time}} & \textbf{v. result} \\ \hline
1.0                                                                                 & 1102                               & 0.099                                  & \multicolumn{1}{c|}{0.370}            & TRUE               & \multicolumn{1}{c|}{0.359}            & TRUE               & \multicolumn{1}{c|}{0.810}            & FALSE              \\ \hline
0.5                                                                                 & 2209                               & 0.216                                  & \multicolumn{1}{c|}{1.772}            & TRUE               & \multicolumn{1}{c|}{6.709}            & FALSE              & \multicolumn{1}{c|}{6.768}            & FALSE              \\ \hline
0.2                                                                                 & 5602                               & 0.801                                  & \multicolumn{1}{c|}{17.243}           & TRUE               & \multicolumn{1}{c|}{16.651}           & TRUE               & \multicolumn{1}{c|}{166.987}          & FALSE              \\ \hline
0.1                                                                                 & 11210                              & 2.388                                  & \multicolumn{1}{c|}{1978.696}                 & TRUE                    & \multicolumn{1}{c|}{3659.402}                 & FALSE                   & \multicolumn{1}{c|}{3649.928}                 & FALSE                   \\ \hline
\end{tabular}
\caption{Model B verification results for Property 2 at varying time precisions.}
\label{tab:modelB_prop2}
\end{table}

Results for Model B are summarized in Table~\ref{tab:modelB_prop2}. For each time precision, we report the number of states, model generation time, verification time, and verification result for three probability thresholds $p=0.25$, $p=0.5$, and $p=0.9$. As we can see, verification times increase with finer time precision, and the property is verified as TRUE only for lower probability thresholds.

\paragraph{Model C Results}

\begin{table}[]
\begin{tabular}{|c|c|c|cc|cc|cc|}
\hline
\multirow{2}{*}{\textbf{\begin{tabular}[c]{@{}c@{}}time \\ precision\end{tabular}}} & \multirow{2}{*}{\textbf{\#states}} & \multirow{2}{*}{\textbf{\begin{tabular}[c]{@{}c@{}}m. gen. \\ time\end{tabular}}} & \multicolumn{2}{c|}{\textbf{p=0.25}}                       & \multicolumn{2}{c|}{\textbf{p=0.5}}                        & \multicolumn{2}{c|}{\textbf{p=0.9}}                        \\ \cline{4-9} 
                                                                                    &                                    &                                        & \multicolumn{1}{c|}{\textbf{v. time}} & \textbf{v. result} & \multicolumn{1}{c|}{\textbf{v. time}} & \textbf{v. result} & \multicolumn{1}{c|}{\textbf{v. time}} & \textbf{v. result} \\ \hline
1.0                                                                                 & 1102                               & 0.099                                  & \multicolumn{1}{c|}{1.437}            & TRUE               & \multicolumn{1}{c|}{1.593}            & FALSE               & \multicolumn{1}{c|}{1.952}            & FALSE              \\ \hline
0.5                                                                                 & 2209                               & 0.216                                  & \multicolumn{1}{c|}{2.963}            & TRUE               & \multicolumn{1}{c|}{9.342}            & FALSE              & \multicolumn{1}{c|}{9.499}            & FALSE              \\ \hline
0.2                                                                                 & 5602                               & 0.801                                  & \multicolumn{1}{c|}{26.483}           & TRUE               & \multicolumn{1}{c|}{28.592}           & FALSE               & \multicolumn{1}{c|}{304.063}          & FALSE              \\ \hline
0.1                                                                                 & 11210                              & 2.388                                  & \multicolumn{1}{c|}{2673.982}                 & TRUE                    & \multicolumn{1}{c|}{4203.124}                 & FALSE                   & \multicolumn{1}{c|}{4381.327}                 & FALSE                   \\ \hline
\end{tabular}
\caption{Model C verification results for Property 2 at varying time precisions.}
\label{tab:modelC_prop2}
\end{table}

Results for Model C are summarized in Table~\ref{tab:modelC_prop2}. For each time precision, we report the number of states, model generation time, verification time, and verification result for three probability thresholds $p=0.25$, $p=0.5$, and $p=0.9$. As we can see, verification times increase with finer time precision, and the property is verified as TRUE only for lower probability thresholds.

\subsection{Property 3: Finish in Good or Disengaged State}

For model A we verify that the rocket will finish the mission in either a good state or the engine will be disengaged:

$\varphi_3 = \langle\langle \text{Rocket} \rangle\rangle \mathbf{G} (\text{Finish} \implies (\text{GoodState} \lor \text{Disengaged}))$

For models B and C we verify that the rocket will finish the mission in either a good state or the engine will be disengaged with at least probability $p$:

$\varphi_{3p} = \langle\langle \text{Rocket} \rangle\rangle^{\geq p} \mathbf{G} (\text{Finish} \implies (\text{GoodState} \lor \text{Disengaged}))$

\paragraph{Model A Results}

\begin{table}[h]
\begin{tabular}{|c|c|c|c|c|}
\hline
\textbf{time precision} & \textbf{\#states} & \textbf{model gen. time} & \textbf{verif. time} & \textbf{verif. result} \\ \hline
1.0                     & 1102              & 0.102                    & 0.013                & TRUE                   \\ \hline
0.5                     & 2209              & 0.213                    & 0.001                & TRUE                   \\ \hline
0.2                     & 5602              & 0.801                    & 0.006                & TRUE                   \\ \hline
0.1                     & 11210             & 2.370                    & 0.008                & TRUE                   \\ \hline
\end{tabular}
\caption{Model A verification results for Property 3 at varying time precisions.}
\label{tab:modelA_prop3}
\end{table}

Results for Model A are summarized in Table~\ref{tab:modelA_prop3}. The model generation time, verification time, and verification result (TRUE/FALSE) are reported for each time precision. The formula was verified as TRUE in all cases.

\paragraph{Model B Results}

\begin{table}[]
\begin{tabular}{|c|c|c|cc|cc|cc|}
\hline
\multirow{2}{*}{\textbf{\begin{tabular}[c]{@{}c@{}}time \\ precision\end{tabular}}} & \multirow{2}{*}{\textbf{\#states}} & \multirow{2}{*}{\textbf{\begin{tabular}[c]{@{}c@{}}m. gen. \\ time\end{tabular}}} & \multicolumn{2}{c|}{\textbf{p=0.25}}                       & \multicolumn{2}{c|}{\textbf{p=0.5}}                        & \multicolumn{2}{c|}{\textbf{p=0.9}}                        \\ \cline{4-9} 
                                                                                    &                                    &                                        & \multicolumn{1}{c|}{\textbf{v. time}} & \textbf{v. result} & \multicolumn{1}{c|}{\textbf{v. time}} & \textbf{v. result} & \multicolumn{1}{c|}{\textbf{v. time}} & \textbf{v. result} \\ \hline
1.0                                                                                 & 1102                               & 0.099                                  & \multicolumn{1}{c|}{0.208}            & TRUE               & \multicolumn{1}{c|}{0.215}            & TRUE               & \multicolumn{1}{c|}{0.218}            & TRUE               \\ \hline
0.5                                                                                 & 2209                               & 0.216                                  & \multicolumn{1}{c|}{1.015}            & TRUE               & \multicolumn{1}{c|}{1.049}            & TRUE               & \multicolumn{1}{c|}{1.392}           & TRUE               \\ \hline
0.2                                                                                 & 5602                               & 0.801                                  & \multicolumn{1}{c|}{6.586}            & TRUE               & \multicolumn{1}{c|}{6.642}            & TRUE               & \multicolumn{1}{c|}{6.543}            & TRUE               \\ \hline
0.1                                                                                 & 11210                              & 2.388                                  & \multicolumn{1}{c|}{29.146}           & TRUE               & \multicolumn{1}{c|}{29.441}           & TRUE               & \multicolumn{1}{c|}{29.758}           & TRUE               \\ \hline
\end{tabular}
\caption{Model B verification results for Property 3 at varying time precisions.}
\label{tab:modelB_prop3}
\end{table}

Results for Model B are summarized in Table~\ref{tab:modelB_prop3}. For each time precision, we report the number of states, model generation time, verification time, and verification result for three probability thresholds $p=0.25$, $p=0.5$, and $p=0.9$. As we can see, the property is verified as TRUE in all cases.

\paragraph{Model C Results}

\begin{table}[]
\begin{tabular}{|c|c|c|cc|cc|cc|}
\hline
\multirow{2}{*}{\textbf{\begin{tabular}[c]{@{}c@{}}time \\ precision\end{tabular}}} & \multirow{2}{*}{\textbf{\#states}} & \multirow{2}{*}{\textbf{\begin{tabular}[c]{@{}c@{}}m. gen. \\ time\end{tabular}}} & \multicolumn{2}{c|}{\textbf{p=0.25}}                       & \multicolumn{2}{c|}{\textbf{p=0.5}}                        & \multicolumn{2}{c|}{\textbf{p=0.9}}                        \\ \cline{4-9} 
                                                                                    &                                    &                                        & \multicolumn{1}{c|}{\textbf{v. time}} & \textbf{v. result} & \multicolumn{1}{c|}{\textbf{v. time}} & \textbf{v. result} & \multicolumn{1}{c|}{\textbf{v. time}} & \textbf{v. result} \\ \hline
1.0                                                                                 & 1102                               & 0.099                                  & \multicolumn{1}{c|}{0.881}            & TRUE               & \multicolumn{1}{c|}{0.893}            & TRUE               & \multicolumn{1}{c|}{0.899}            & FALSE               \\ \hline
0.5                                                                                 & 2209                               & 0.216                                  & \multicolumn{1}{c|}{2.594}            & TRUE               & \multicolumn{1}{c|}{2.690}            & TRUE               & \multicolumn{1}{c|}{3.527}           & FALSE               \\ \hline
0.2                                                                                 & 5602                               & 0.801                                  & \multicolumn{1}{c|}{9.720}            & TRUE               & \multicolumn{1}{c|}{10.351}            & TRUE               & \multicolumn{1}{c|}{10.550}            & FALSE               \\ \hline
0.1                                                                                 & 11210                              & 2.388                                  & \multicolumn{1}{c|}{42.038}           & TRUE               & \multicolumn{1}{c|}{43.403}           & TRUE               & \multicolumn{1}{c|}{43.602}           & FALSE               \\ \hline
\end{tabular}
\caption{Model C verification results for Property 3 at varying time precisions.}
\label{tab:modelC_prop3}
\end{table}

Results for Model C are summarized in Table~\ref{tab:modelC_prop3}. For each time precision, we report the number of states, model generation time, verification time, and verification result for three probability thresholds $p=0.25$, $p=0.5$, and $p=0.9$. As we can see, the property is verified as TRUE only for lower probability thresholds.

\section{Conclusion and Future Work}\label{sec:conclusion}

This work has presented a comprehensive framework for formal verification of probabilistic multi-agent systems applied to ballistic rocket flight trajectories using Probabilistic Alternating-Time Temporal Logic (PATL). Our research bridges the gap between theoretical formal verification methods and practical aerospace engineering challenges, demonstrating how strategic reasoning can enhance safety assurance in critical systems.

The key contributions of this work include:

\begin{itemize}
    \item Development of an innovative verification framework specifically designed for analyzing safety properties of ballistic rockets engineered to achieve microgravity conditions for scientific experimentation.
    
    \item Integration of authentic flight telemetry data with high-fidelity simulation to construct probabilistic state transition systems that rigorously account for environmental stochasticity, particularly meteorological variability.
    
    \item Formalization of mission-critical safety properties through PATL specifications that enable systematic identification of trajectory deviation states where rockets risk landing in prohibited or hazardous zones.
    
    \item Implementation of a practical verification approach validated against the Perun Rocket system developed by SpaceForest, demonstrating real-world applicability of advanced formal methods.
    
    \item Comparative analysis of three distinct modeling approaches (non-probabilistic, probabilistic with perfect information, and probabilistic with imperfect information) that reveals important trade-offs between model fidelity and verification feasibility.
\end{itemize}

Our experimental validation demonstrates that while deterministic models (Model A) consistently verify safety properties across all time precisions, probabilistic models reveal critical limitations when higher confidence thresholds are required. These findings underscore the importance of appropriate confidence threshold selection when deploying formal verification in safety-critical aerospace applications. The verification results also highlight the computational challenges inherent in strategic ability verification for probabilistic systems with imperfect information. As time precision increases, both state space complexity and verification time grow significantly. This demonstrates the practical limitations of current model checking approaches when applied to high-fidelity aerospace systems.

Future work will focus on several promising directions. First of all, in order to reduce the state space explosion problem, we plan to develop efficient model reduction techniques specifically tailored for aerospace applications. Additionally, we aim to explore new verification algorithms that can better handle the complexities of probabilistic multi-agent systems with imperfect information such as fix-point approximations.

This research establishes a foundation for applying formal verification methods to commercial spaceflight systems, potentially setting new standards for safety assurance in the rapidly growing private space industry. By demonstrating the practical utility of PATL in analyzing real-world aerospace systems, we hope to encourage wider adoption of formal methods throughout the space technology sector, ultimately contributing to safer and more reliable access to space for scientific research and commercial applications.

\para{Acknowledgements}
The work has been supported by NCBR Poland and FNR Luxembourg under the PolLux/FNR-CORE project SpaceVote (POLLUX-XI/14/SpaceVote/2023 and C22/IS/17232062/SpaceVote).
For the purpose of open access, and in fulfilment of the grant agreement, the authors have applied CC BY 4.0 license to any Author Accepted Manuscript version arising from this submission.

\bibliographystyle{plain}
\bibliography{wojtek,wojtek-own}

\end{document}